# Thickness dependence study of current-driven ferromagnetic resonance in $Y_3Fe_5O_{12}$/heavy metal bilayers

Z. Fang,[1,*] A. Mitra,[2] A. L. Westerman,[2] M. Ali,[2] C. Ciccarelli,[1] O. Cespedes,[2] B. J. Hickey[2] and A. J. Ferguson[1]

[1]*Microelectronics Group, Cavendish Laboratory, University of Cambridge, JJ Thomson Avenue, Cambridge CB3 0HE, United Kingdom.*

[2]*School of Physics and Astronomy, University of Leeds, Leeds LS2 9JT, United Kingdom.*

We use ferromagnetic resonance to study the current-induced torques in YIG/heavy metal bilayers. YIG samples with thickness varying from 14.8 nm to 80 nm, with Pt or Ta thin film on top, are measured by applying a microwave current into the heavy metals and measuring the longitudinal DC voltage generated by both spin rectification and spin pumping. From a symmetry analysis of the FMR lineshape and its dependence on YIG thickness, we deduce that the Oersted field dominates over spin-transfer torque in driving magnetization dynamics.

*Introduction* — Insulating magnetic materials have recently played an important role in spintronics, since they allow pure spin currents to flow without associated charge transport. Within the family of ferromagnetic insulators, yttrium iron garnet (YIG) holds a special place owing to several favourable properties, including ultra-low damping, high Curie temperature and chemical stability [1–3]. By growing an overlayer of heavy metal (HM), such as platinum or tantalum, several important spintronic phenomena have been explored in the YIG/HM bilayer system, including the magnetic proximity effect [4,5], spin pumping [6,7], spin Hall magnetoresistance (SMR) [8,9], spin Seebeck effect [10,11] and so on. Furthermore, the spin Hall effect in HM can convert a charge current into a transverse pure spin current, making it possible to manipulate the ferromagnetic insulator by spin transfer torque (STT). Recently, several groups have reported controlling the damping in YIG by applying a DC charge current in a Pt capping layer [12], by which spin-Hall auto-oscillation can be realized [13,14]. Replacing the DC current with a microwave current, the electrical signal in Pt can also be transmitted via spin waves in YIG [3]. In order to further explore the application of the YIG/HM system, it is necessary to understand the torque on YIG induced by the charge current in HM.

Current-induced ferromagnetic resonance (CI-FMR) is an effective method to characterize ferromagnetic samples at micrometre-scale and quantify the current-induced magnetic torques in ferromagnetic/HM bilayer systems [15]. As shown in FIG. 1(a), an oscillating charge current in the HM layer generates a perpendicular pure spin-current oscillating at the same frequency via the spin Hall effect. This oscillating spin current flows into the ferromagnetic layer, exerting an oscillating STT, which can drive magnetization precession when the FMR condition is satisfied [15–19]. Since no charge current is required to flow inside the ferromagnetic layer in this process, it is possible to extend this method to ferromagnetic insulator/HM bilayers. Instead of penetrating into the ferromagnetic layer, the electrons undergo spin-dependent scattering at the interface between the ferromagnetic insulator and the HM, transferring angular momentum to the ferromagnetic insulator. The accompanied Oersted field generated by the charge current in Pt can also drive the magnetization precession. Although from the symmetry point of view, the torques induced by the Oersted field and the field-like component of STT are indistinguishable from each other, in our work we confirm that the driving field is dominated by the Oersted contribution by repeating the measurement with Pt and Ta.

[*] zf231@cam.ac.uk

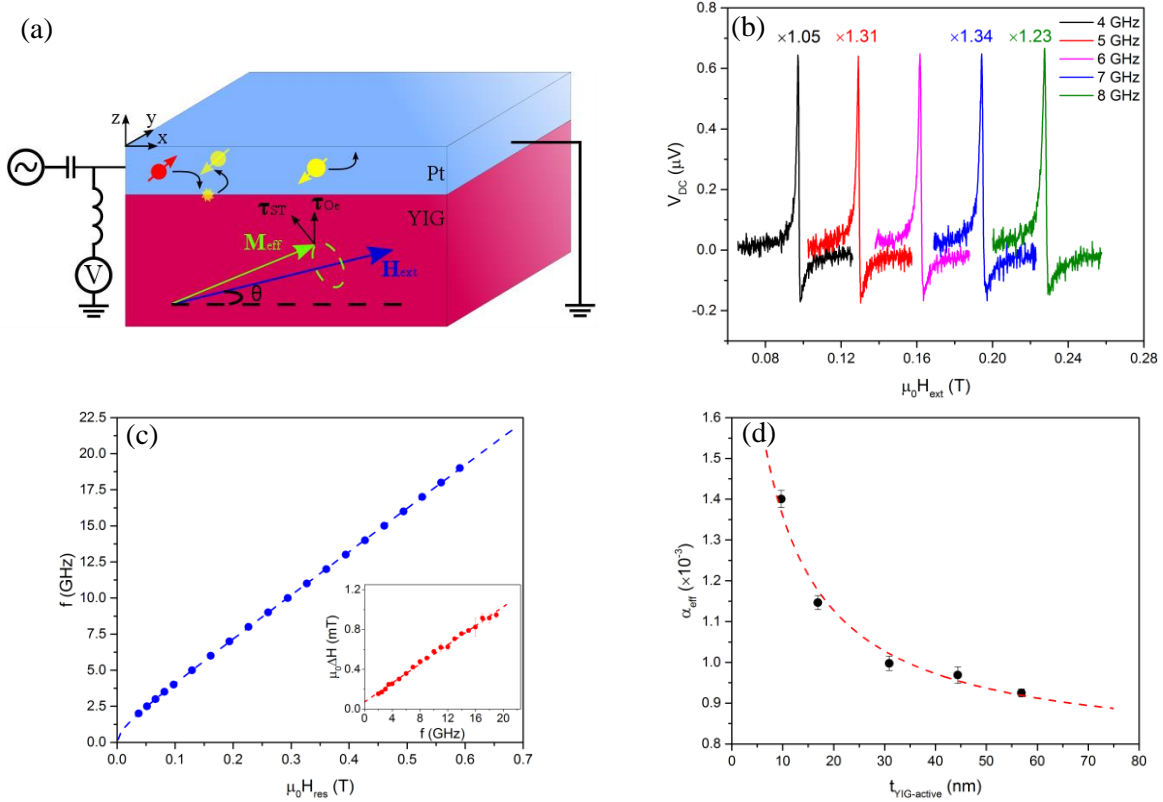

**FIG. 1.** (a) Scheme of CI-FMR in YIG/Pt and the experiment setup. (b)-(c) Results from a YIG(14.8)/Pt(4.2) sample: (b) spectra of CI-FMR at 4-8 GHz; (c) resonance frequency $f$ as a function of the resonant field $\mu_0 H_{res}$, fitted with the in-plane Kittel's formula in dashed line. Inset: frequency dependence of the FMR linewidth $\mu_0 \Delta H$. (d) Plot of the effective damping $\alpha_{eff}$ as a function of $t_{YIG\text{-}active}$. Red dashed line represents the fitting result using EQ. (4).

Recently, several groups have studied the CI-FMR in YIG/Pt both theoretically [20,21] and experimentally [22–24]. Using the theoretical model built by Chiba et al. [21], Schreier et al. did the first experiment on in-plane CI-FMR in YIG/Pt and identified the current-induced torque by the symmetry and the lineshape of the signals [22]. Sklenar et al. then repeated the experiment for an out-of-plane external magnetic field [23]. Very recently, Jungfleisch et al. imaged the current-driven magnetization precession in YIG/Pt at resonance condition with Brillouin light scattering spectroscopy and argued that uniform precession is no longer applicable at high microwave power [24]. The behaviour of CI-FMR in YIG/HM, however, should depend on the thickness of the films [20], which is one aspect that remains under explored to the best of our knowledge.

Here, we study the current-induced torque in YIG/Pt (or Ta) bilayer structures with different YIG thickness using CI-FMR. By applying a microwave current into the HM and sweeping the external magnetic field in the plane of the devices, a DC voltage is observed at the resonance condition. This DC voltage is generated simultaneously by two mechanisms [20]: spin rectification and spin pumping. The nature of the torque can be understood from the lineshape and the symmetry of the DC voltage obtained from different samples.

*Samples and experimental methods* — YIG films with different thickness (listed in TAB. 1) were grown using RF sputtering on substrates of (111) gadolinium gallium garnet at a pressure of 2.4 mTorr. Since the deposited YIG was nonmagnetic, the film was annealed *ex situ* at 850°C for 2 hours. A layer of 4.2 ± 0.1 nm Pt (or 5.0 ± 0.1 nm Ta) was then deposited via DC magnetron sputtering. Both YIG and Pt thicknesses were measured by x-ray reflectivity. The samples were patterned into 5×50 μm² bars by using optical lithography and argon ion milling. After a second round of optical lithography, a layer of 5 nm Cr/50 nm Au were evaporated as the contact electrodes. Each bar was mounted on a low-loss dielectric circuit board and connected to a microstrip transmission line via wire bonding. By using a bias-tee, the DC voltage across the bar was measured at the same time as microwave power was applied. A magnetic field $\mathbf{H}_{ext}$ was swept in the film plane at an angle $\theta$ with respect to the bar, as defined in FIG. 1(a).

**TAB. 1.** Summary of sample characteristics. The HM cap is Pt unless specified.

| $t_{YIG}$ (nm) | SMR ($10^{-5}$) | $\gamma_{eff}/2\pi$ (GHz/T) | $M_{eff}$ (kA/m) | $\alpha_{eff}$ ($10^{-3}$) | $K_{2\perp}$ (kJ/m³) |
|---|---|---|---|---|---|
| 14.8 | 4.8 ± 0.6 | 30.0 ± 0.1 | 69 ± 3 | 1.41 ± 0.02 | 12.6 ± 1.3 |
| 22 | 5.9 ± 1.0 | 29.9 ± 0.1 | 81 ± 4 | 1.15 ± 0.03 | 11.2 ± 1.1 |
| 36 | 2.9 ± 0.3 | 29.8 ± 0.1 | 82 ± 3 | 1.00 ± 0.02 | 11.1 ± 1.0 |
| 49.5 | 5.1 ± 0.1 | 29.8 ± 0.1 | 82 ± 3 | 0.97 ± 0.02 | 11.1 ± 1.0 |
| 62 | 3.1 ± 0.5 | 29.9 ± 0.1 | 77 ± 4 | 0.93 ± 0.04 | 11.6 ± 1.2 |
| 80 (Ta) | 1.2 ± 0.5 | 28.1 ± 0.9 | 90 ± 7 | 1.36 ± 0.31 | 10.2 ± 0.9 |

*Signal symmetry* — The lineshape and the symmetry of the resonance in the DC voltage that we measure depends both on how the magnetization precession is driven and how the DC voltage is generated. As for the driving mechanism, when a microwave current $I_0 e^{j\omega t}$ flows though the HM, two types of torques are expected to act on YIG: a field-like torque $\boldsymbol{\tau}_{Oe} = \mathbf{M} \times \mathbf{h}_{Oe}$ induced by the Oersted field $\mathbf{h}_{Oe} // \mathbf{y}$, where $\mathbf{y}$ is the unit vector along the y axis, and an antidamping-like STT $\boldsymbol{\tau}_{ST} = \mathbf{M} \times \mathbf{h}_{ST}$ induced by an effective field $\mathbf{h}_{ST} // \mathbf{y} \times \mathbf{M}$. If $\mathbf{M}$ is in the x-y plane, both torques reach their maximum when $\mathbf{M}$ is along the x-axis, and become zero when $\mathbf{M}$ is perpendicular to the current direction.

As for the generation of the longitudinal voltage, two mechanisms are mainly involved: spin rectification and spin pumping. At the FMR condition, the oscillating magnetization leads to a time-dependent SMR in the HM at the same frequency: $R = R_0 + \Delta R \cos^2\theta(t)$ [8,25], which rectifies the microwave current, inducing a DC voltage along the bar. We have characterised the SMR for each YIG thickness by measuring the resistance of the Pt bar as an external in-plane magnetic field is rotated. The results are reported in TAB. 1. The spin-rectification DC voltage consists of a symmetric ($V_{sym\text{-}SR}$) and an antisymmetric ($V_{asy\text{-}SR}$) Lorentzian components [26,27]:

$$V_{DC} = V_{sym\text{-}SR} \frac{\Delta H^2}{\left(H_{ext} - H_{res}\right)^2 + \Delta H^2} + V_{asy\text{-}SR} \frac{\Delta H \left(H_{ext} - H_{res}\right)}{\left(H_{ext} - H_{res}\right)^2 + \Delta H^2} \quad (1)$$

$$V_{sym\text{-}SR} = \frac{I_0 \Delta R}{2} \frac{\sqrt{H_{res}\left(H_{res} + M_{eff}\right)}}{\Delta H \left(2H_{res} + M_{eff}\right)} h_{ST} \sin 2\theta \quad (2)$$

$$V_{asy\text{-}SR} = \frac{I_0 \Delta R}{2} \frac{\left(H_{res} + M_{eff}\right)}{\Delta H \left(2H_{res} + M_{eff}\right)} h_{Oe} \sin 2\theta \cos\theta \quad (3)$$

Here, $H_{ext}$, $H_{res}$ and $\Delta H$ are external applied magnetic field, the resonant field and the linewidth (half width at half maximum) respectively; $M_{eff} = M_s - 2K_{2\perp}/\mu_0 M_s$ is the effective magnetization, where $M_s$, $K_{2\perp}$ and $\mu_0$ refer to the saturation magnetization, the interface anisotropy energy density and the vacuum permeability respectively. EQ. (2)-(3) show that $V_{\text{sym-SR}}$ is induced by the out-of-plane field $\mathbf{h}_{\text{ST}}$ while $V_{\text{asy-SR}}$ is induced by the Oersted field $\mathbf{h}_{\text{Oe}}$.

The other mechanism that could lead to a DC longitudinal voltage is spin pumping. The DC voltage from spin pumping, irrespective of the driving mechanism, is described by a symmetric Lorentzian component $V_{\text{sym-SP}}$, being independent of the phase between the microwave current in the HM and the precessing magnetization. TAB. 2 summarizes the lineshape and the angle dependence of the DC voltage for each of the driving and detecting mechanisms discussed above. Here, the effective damping factor $\alpha_{\text{eff}}$ is a function of $t_{\text{YIG}}$ and includes the spin pumping term $\alpha_{\text{SP}}$ [28]:

$$\alpha_{\text{eff}}(t_{\text{YIG}}) = \alpha_0 + \alpha_{\text{SP}} = \alpha_0 + \frac{g\mu_B}{4\pi M_s t_{\text{YIG}}} g_{\text{eff}}^{\uparrow\downarrow} \qquad (4)$$

where $\alpha_0$ is the intrinsic Gilbert damping coefficient of YIG without HM cap respectively; $g$ is the g-factor; $\mu_B$ is the Bohr magneton; $g_{\text{eff}}^{\uparrow\downarrow}$ is the interface effective spin mixing conductance taking into account the backflow. Assuming that $\alpha_0$ does not change with $t_{\text{YIG}}$, $g_{\text{eff}}^{\uparrow\downarrow}$ can be determined by measuring $\alpha_{\text{eff}}$ for samples of different YIG thickness.

**TAB. 2.** Summary of resonance DC signal components involved, with their Lorentzian lineshape and dependence on $\theta$, spin Hall angle $\vartheta_{\text{SH}}$, $t_{\text{YIG}}$ and $\alpha_{\text{eff}}$. $C_i$ are the positive coefficients independent from the parameters listed above.

| Driving | Detecting | lineshape | Dependence on $\theta$, $\vartheta_{\text{SH}}$, $t_{\text{YIG}}$, and $\alpha_{\text{eff}}$ |
|---|---|---|---|
| $\mathbf{h}_{\text{ST}}$ | SR | Symmetric | $-C_{\text{ST-SR}}\left[\vartheta_{\text{SH}}^3/(\alpha_{\text{eff}} t_{\text{YIG}})\right]\sin 2\theta \cos\theta$ |
| | SP | Symmetric | $C_{\text{ST-SP}}\left[\vartheta_{\text{SH}}^3/(\alpha_{\text{eff}} t_{\text{YIG}})^2\right]\sin 2\theta \cos\theta$ |
| $\mathbf{h}_{\text{Oe}}$ | SR | Anti-symmetric | $-C_{\text{Oe-SR}}\left(\vartheta_{\text{SH}}^2/\alpha_{\text{eff}}\right)\sin 2\theta \cos\theta$ |
| | SP | Symmetric | $C_{\text{Oe-SP}}\left(\vartheta_{\text{SH}}/\alpha_{\text{eff}}^2\right)\sin 2\theta \cos\theta$ |

*Results and Discussion* — FIG. 1(b) shows an example of CI-FMR signals measured at $f$ = 4-8 GHz and $\theta = 45°$ for a sample YIG(14.8)/Pt(4.2). The resonances are well described by a Lorentzian lineshape consisting of symmetric and antisymmetric components. FIG. 1(c) plots the frequency dependence of resonance field and linewidth (inset), which are well fitted by the in-plane Kittel formula $f = (\mu_0\gamma_{\text{eff}}/2\pi)[H_{\text{res}}(H_{\text{res}} + M_{\text{eff}})]^{1/2}$ and the linear linewidth function $\mu_0\Delta H = \mu_0\Delta H_0 + 2\pi f\alpha_{\text{eff}}/\gamma_{\text{eff}}$ respectively. Here, $\gamma_{\text{eff}}$ is the effective gyromagnetic ratio and $\Delta H_0$ is the inhomogeneous linewidth broadening. From the fitting we calculate the parameters, $\gamma_{\text{eff}}$, $M_{\text{eff}}$ and $\alpha_{\text{eff}}$, as summarized in TAB. 1, together with those obtained from each of the samples considered in this study. By using a vibrating sample magnetometer (VSM), we measured the values of $M_s$ to be 180 ± 20 kA/m for all the samples at room temperature, and $K_{2\perp}$ can now be calculated (TAB. 1). From VSM measurement, we also find that there is a 5.1 ± 0.1 nm-thick non-magnetic dead layer in our YIG films. Therefore, we define the thickness of active YIG layer as $t_{\text{YIG-active}} = t_{\text{YIG}} - 5.1$ in nm, and this value should be used in our calculation. As shown in FIG. 1(d), the value of $\alpha_{\text{eff}}$ for different $t_{\text{YIG-active}}$ is well fitted by EQ. (4) and we find the values of $\alpha_0$ and $g_{\text{eff}}^{\uparrow\downarrow}$ to be (8.1 ± 0.1) × 10$^{-4}$ and (7.1 ± 0.2) × 10$^{17}$ m$^{-2}$ respectively, in reasonably good agreement with the literature [29].

To characterize the current-induced torque, we now analyse the angle dependence of the symmetric and the antisymmetric components of the resonance signal. FIG. 2(a) shows the result obtained from the YIG(14.8)/Pt(4.2) sample. The $V_{asy}$ is fitted well with a $-\sin2\theta\cos\theta$ function alone (red dash), in agreement with a resonance driven by the Oersted field and detected by spin-rectification (TAB. 2).) In contrast, $V_{sym}$ is fitted by the sum of a $\sin2\theta\cos\theta$ term (orange dash) and a $\sin\theta$ (green dash) term, noted as $V_{sym\text{-}sin2\theta cos\theta}$ and $V_{sym\text{-}sin\theta}$ respectively. All components are linear in power (FIG. 2(b)), indicating that the small-angle precession approximation is satisfied. By carrying a quantitative analysis of $V_{asy}$ based on EQ. (3), we extract the value of the effective field that generates the torque for each sample (FIG. 2(c)), normalized to a unit current density of $j_c = 10^{10}$ A/m$^2$. This can be compared with the value of the Oersted field calculated from Ampere's law as $\mu_0 h_{Oe} = \mu_0 j_c t_{Pt}/2 \approx 26$ µT (red dash in FIG. 2(c)), where $t_{Pt}$ is the thickness of Pt. The good agreement between the two values confirms that the field-like torque is mainly attributed to the Oersted field.

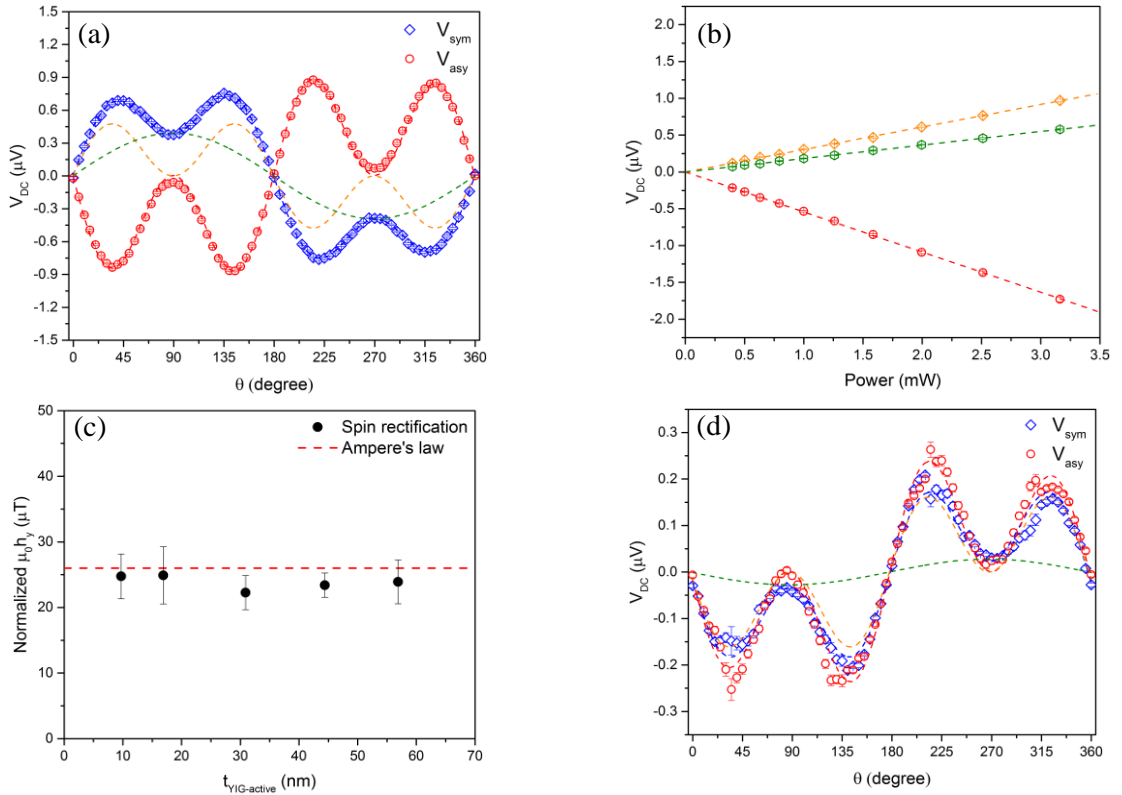

**FIG. 2.** (a) Angle dependence of the symmetric part $V_{sym}$ (blue) and anti-symmetric part $V_{asy}$ (red) from YIG(14.8)/Pt(4.2) at 8 GHz. Dashed lines are fitting results, where $V_{asy}$ is fitted by $\sin2\theta\cos\theta$, while $V_{sym}$ needs a $\sin\theta$ term (green) in addition to the $\sin2\theta\cos\theta$ term (orange). (b) Power dependence of the three resonance components at 8 GHz. (c) Oersted field $\mu_0 h_y$ calculated from Ampere's law (red dashed line) and $V_{asy}$ using EQ. (3) (black dot) for each sample, normalized to $j_c = 10^{10}$ A/m. (d) Angle dependence measurement from a YIG(80)/Ta(5.0) sample.

The analysis of the $\sin2\theta\cos\theta$ term (orange dash) in symmetric component is richer, since it contains three different terms as shown in TAB. 2. Despite this, we can still identify the main driving mechanism by comparing $V_{sym\text{-}sin2\theta cos\theta}$ to $V_{asy}$. FIG. 3 plots the ratio $V_{sym\text{-}sin2\theta cos\theta}/V_{asy}$ in each sample with respect to $1/\alpha_{eff}$, showing a linear dependence. Referring to TAB. 2, only $|V_{Oe\text{-}SP}/V_{Oe\text{-}SR}| \propto 1/\alpha_{eff}$, while the ratios between other terms have a more complicated relation to $\alpha_{eff}$, as $\alpha_{eff}$ also depends on $t_{YIG\text{-}active}$ (EQ. (4)). From this we conclude that $V_{sym\text{-}sin2\theta cos\theta}$ can be mainly attributed to the spin pumping driven by the Oersted field. In addition to this, we carried out an experiment in which we replaced the Pt with Ta. FIG. 2(d) shows the angle dependence for a YIG(80)/Ta(5.0) sample. While $V_{sym}$ changes its sign compared with the YIG/Pt case, the sign of $V_{asy}$ stays the same. The change in the sign of $V_{sym}$ is explained with the opposite sign of the spin-pumping, which results from the opposite value of the spin-Hall angle of Ta compared with Pt [15,16,30]. The fact that the sign of $V_{asy}$ does not depend on the sign of the spin-Hall angle of the metal layer further confirms that the Oersted field dominates over the field-like STT in driving the magnetization dynamics in our samples [31].

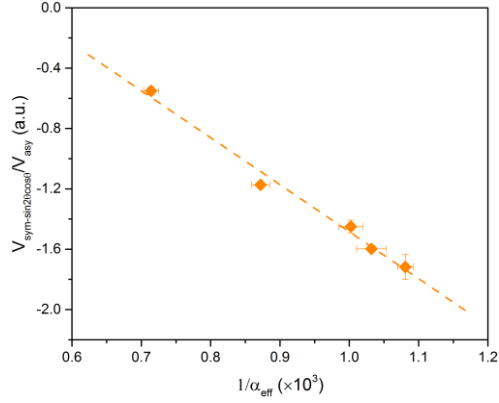

**FIG. 3.** Plot of the ratio $V_{sym\text{-}sin2\theta cos\theta}/V_{asy}$ as a function of $1/\alpha_{eff}$, measured at 8 GHz. The dashed line represents the linear fitting.

We also briefly comment on an additional $\sin\theta$ (green dash) term that appears in the fitting of $V_{sym}$. We note that when measuring other material systems in our setup, e.g. Co/Pt [27] or Py/Pt [32], this $\sin\theta$ component is absent, indicating its origin in the sample. This term was previously attributed to an on-resonance contribution from the longitudinal spin Seebeck effect [24]. However, neither STT nor Oersted field can drive FMR at $\theta = 90°$, where this term is maximum.

*Conclusion* — In conclusion, we have used CI-FMR to investigate the charge-current-induced torque on YIG magnetization in a series of YIG/HM samples with different YIG thickness between 14.8 nm and 80 nm. Our measurements show that the Oersted field gives the dominant contribution to driving the magnetisation precession and should therefore be taken into account when carrying out CI-FMR studies in YIG/HM systems.

**ACKNOWLEDGMENTS**

We thank L. Abdurakhimov for the experiment help, and T. Jungwirth, M. Jungfleisch & A. Hoffmann for the valuable discussions. Z.F. thanks the Cambridge Trusts for financial support. B.J.H. thanks D. Williams and Hitachi Cambridge for support and advice. A.J.F. is supported by ERC grant 648613 and a Hitachi Research Fellowship. The research leading to these results has received funding from the European Union Seventh Framework Programme (FP - People-2012-ITN) under Grant No. 316657 (SpinIcur) and the Engineering and Physical Sciences Research Council (EPSRC).